\pdfoutput=1
\documentclass[conference]{IEEEtran}

\usepackage{amsmath,amsfonts,amssymb}
\usepackage{graphicx}
\usepackage{cite}
\usepackage{multirow}
\usepackage{algorithm}
\usepackage{algorithmic}
\usepackage{subfigure}
\usepackage{multirow}
\usepackage{url}
\usepackage{color}
\usepackage{array}
\usepackage[margin=1in]{geometry}
\usepackage{subfig}
\usepackage{tabularx}

\graphicspath{{./figs/}}

\IEEEoverridecommandlockouts

\newtheorem{definition}{Definition}
\newtheorem{lemma}{Lemma}

\newtheorem{proposition}{Proposition}

\newcommand{\gammatd}{\tilde{\gamma}}

\newcommand{\htd}{\tilde{\mathbf h}}
\newcommand{\Ctd}{\tilde{C}}
\newcommand{\Rtd}{\tilde{R}}

\newcommand{\gammabd}{{\boldmath $\gamma$}}
\newcommand{\gammatdbd}{{\boldmath $\tilde{\gamma}$}}

\begin{document}
\title{Electromagnetic Lens-focusing Antenna Enabled Massive MIMO}
\author{\authorblockN{Yong~Zeng, Rui~Zhang, and Zhi~Ning~Chen\\}
\authorblockA{Department of Electrical and Computer Engineering, National University of Singapore\\
zengyongedu@hotmail.com, elezhang@nus.edu.sg, eleczn@nus.edu.sg
}
}
\IEEEspecialpapernotice{(Invited Paper)}
\maketitle

\begin{abstract}
Massive multiple-input multiple-output (MIMO) techniques have been recently advanced to tremendously improve the performance of wireless networks. 
However, the use of very large antenna arrays brings new issues, such as the significantly increased hardware cost and signal processing cost and complexity. In order to reap the enormous gain of  massive MIMO and yet reduce its cost to an affordable level, this paper proposes  a novel system design by integrating  an electromagnetic (EM) lens with the large antenna array, termed \emph{electromagnetic  lens antenna} (ELA).  An ELA  has the capability of focusing the power  of any incident  plane wave passing through the EM lens  to a small subset of the antenna array, while the location of focal area is dependent on the angle of arrival (AoA) of the wave. As compared to conventional antenna arrays without the EM lens, the proposed system can substantially  reduce the number of required  radio frequency (RF) chains at the receiver and hence, the implementation costs.  In this paper, we  investigate  the proposed system under a simplified single-user uplink transmission setup, by characterizing the power distribution of the ELA as well as the resulting channel model. Furthermore, by assuming  antenna selection used at the receiver, we show the throughput  gains of the proposed  system over conventional antenna arrays given the same number of selected antennas.
\end{abstract}

\section{Introduction}
Multi-antenna or multiple-input multiple-output (MIMO)  systems have been shown to offer great advantages over conventional  single-antenna systems in point-to-point, single-cell multiuser MIMO, as well as multi-cell MIMO transmissions \cite{36,377,130}. Recently, an even more advanced  multi-antenna technique  known as massive MIMO has been proposed, where antenna arrays with  large or ultra-large  number of elements are deployed  at the base stations (BSs) to reap the MIMO transmission benefits on a greater scale (see \cite{374} and references therein). For example, given  a massive MIMO system of $100$ antennas serving about $40$ terminals with  the same time-frequency resource, a simultaneous increase of both the spectral efficiency (in bits/sec/Hz) by $10$ times and the energy efficiency (in bits/Joule) by $100$ times can be achieved \cite{375}, as compared to the reference system with one single antenna serving a single terminal. 
 Other benefits of massive MIMO include the asymptotic optimality of simple linear processing schemes such as maximal-ratio combining (MRC), the resilience against failures of individual antenna elements, and the possibility to simplify the multiple-access techniques, etc.

Despite many  promising benefits,  massive MIMO systems are faced with new challenges, which, if not tackled successfully,  could roadblock their widely deployment in practice.   Firstly, the use of large antenna array increases  the hardware cost considerably. Even with inexpensive  antenna elements, the cost associated with the radio frequency (RF) elements, which include mixers, D/A and A/D converters, and amplifiers etc., grows up significantly  with the increasing number of antennas used. Secondly, the complexity of signal processing increases drastically due to the large number of branch signals that need to be processed from all antennas, as well as the increased number of channel parameters that need to be estimated for coherent communications. Thirdly, the total energy consumption including that for the RF chains may be greatly increased due to the use of large number of antennas, which can  even negate the power saving with massive MIMO transmissions.
\begin{figure}
\centering
\includegraphics[scale=0.55]{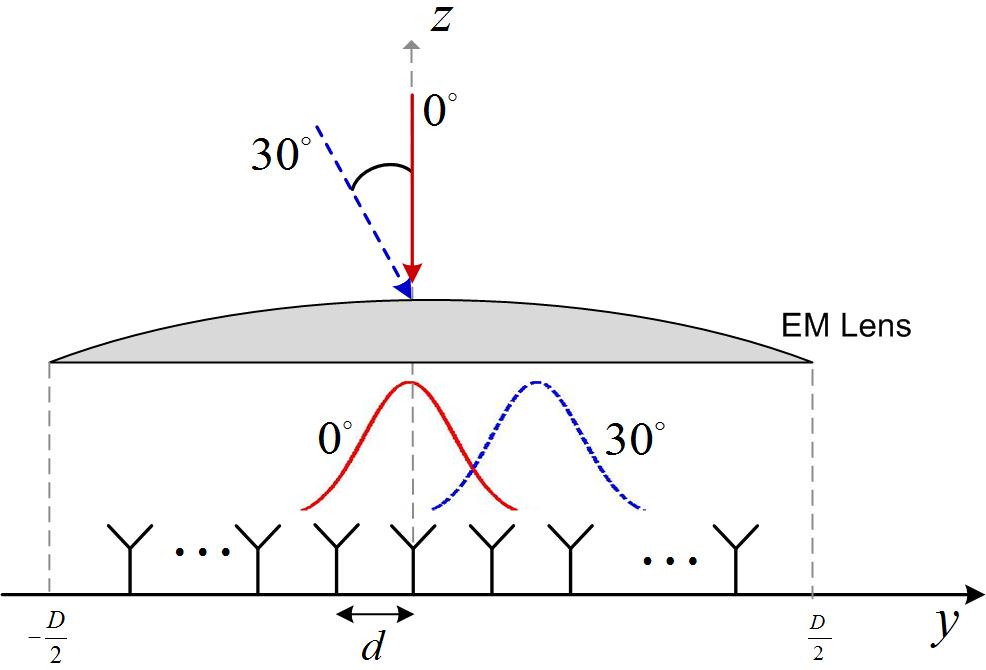}
\caption{Proposed design with the EM-lens  embedded antenna array. \vspace{-4ex}}
\label{F:proposed}
\end{figure}

\begin{figure}%
\centering
\includegraphics[scale=0.23]{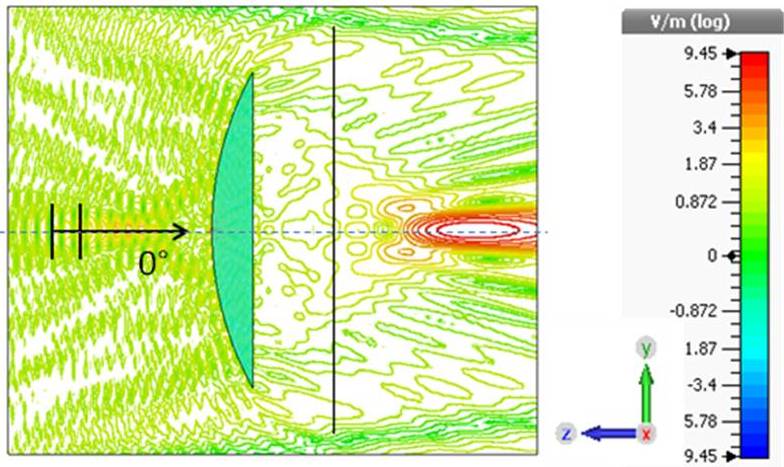}%
\\
(a)
\\
\includegraphics[scale=0.23]{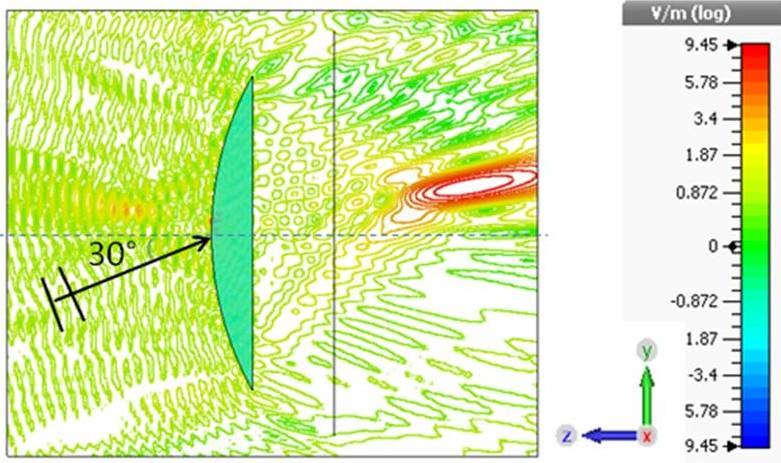}
\\
(b)
\caption{E-field distribution with AoA of (a) $\theta=0^{\circ}$; and (b) $\theta=30^{\circ}$  \cite{376}. \vspace{-4ex}}%
\label{F:RFfileddistribution}%
\end{figure}

In order to capture the promising gains of massive MIMO and yet reduce its cost to an affordable level, we propose in this paper a novel system design as shown in Fig.~\ref{F:proposed}, where a new component called electromagnetic (EM) lens is integrated with the large antenna array, termed {\it electromagnetic lens antenna} (ELA).  An EM lens is usually built with dielectric material with curved front and/or rear surfaces. With the geometry carefully designed, the EM lens is able to change the paths of the incident EM waves in a desired manner so that the arrival signal energy is focused to a much smaller region of the antenna array. Furthermore, for a given EM lens, the spatial power distribution of any uniform plane wave passing through it  is determined by the angle of arrival (AoA) of the incident wave. This is demonstrated in  Fig.~\ref{F:RFfileddistribution}, where the E-field distribution of a practical EM lens with  the  refractive index of two is shown. The  aperture diameter and thickness of the EM lens is $20\lambda$ and $1.6\lambda$, respectively, where $\lambda$ is the wavelength in free space. It is observed that as the incident angle $\theta$ changes from $0^{\circ}$ to $30^{\circ}$, the location of the strongest E-field distribution sweeps  accordingly.

In this paper, for an initial investigation we apply   the proposed ELA system to a simplified single-user uplink communication setup. We first characterize the power distribution of the EM lens along the line where the antenna array is placed. The channel model with the receiver ELA by incorporating  the power distribution  of the EM lens is then established.  Furthermore, we evaluate  the throughput  gains of the proposed system over conventional antenna arrays with the antenna selection (AS) scheme applied  at the receiver. We show that the proposed ELA design can achieve the same capacity as the conventional  antenna array without the EM lens, but requires   significantly reduced number of active antennas, thus yielding much lower energy cost and signal  processing cost and complexity  at the receiver.


\section{System Model}\label{sec:model}
\begin{figure}
\centering
\includegraphics[scale=0.40]{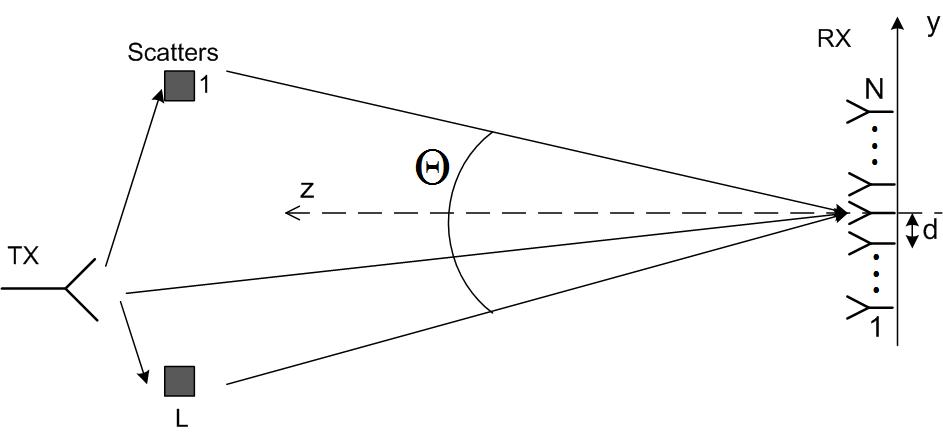}
\caption{SIMO system with linear antenna array at the receiver.}
\label{F:Model}
\end{figure}

 In this paper, we consider a simplified narrowband uplink communication as shown in Fig.~\ref{F:Model}, where a single  mobile user with one omni-directional antenna transmits to the BS with a large uniform linear array (ULA) consisting of $N\gg 1$ antenna elements deployed along the y-axis and separated by distance of $d$. Without loss of generality, we assume that the antenna array is centered at $y=0$, so that the location $y_i$ of the $i$th element is
 \begin{align}\label{eq:yi}
 y_i=-\frac{(N-1)d}{2}+(i-1)d, \ i=1,\cdots, N.
 \end{align}

\begin{figure}
\centering
\includegraphics[scale=0.5]{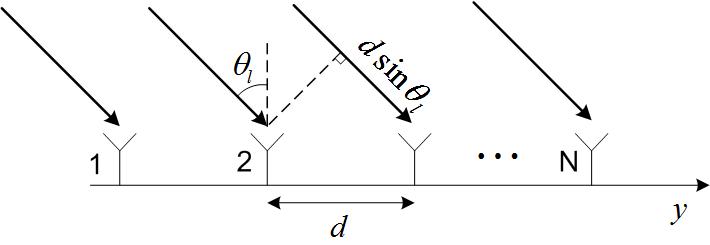}
\caption{The $l${th} incident plane wave with AoA of  $\theta_l$.\vspace{-3ex}}
\label{F:UPW}
\end{figure}

We assume that the transmitted signal from the user  arrives at the BS antenna array via $L\geq 1$ paths, where the $l${th} path, $l=1,\cdots,L$, impinges as a plane wave with AoA of $\theta_l$, as shown in Fig.~\ref{F:UPW}. With plane wave assumption, the surfaces of constant phase of  an incident wave are parallel planes normal to the direction of incidence, and this is justified when the antenna aperture of the ULA is much smaller than the transmission distance between the  user and BS. We further denote  the angular spread of the incident waves by $\Theta$, where $\Theta\leq \pi$; as a result, we have $\theta_l\in \left[-\frac{\Theta}{2}, \frac{\Theta}{2}\right], \forall l$. The channel coefficient $h_i$ between the user and the $i${th} antenna element at the BS is then represented as (without the EM lens applied  yet)
 \begin{align}\label{eq:hn}
  h_i=\sum_{l=1}^L \sqrt{g_l} e^{j(\alpha_l+\frac{2\pi}{\lambda}i d \sin \theta_l)},\ i=1,\cdots, N,
 \end{align}
 where $\lambda$ denotes the wavelength, $j$ is the imaginary unit with $j^2=-1$, $g_l$ represents the power gain of the $l${th} component, and $\alpha_l$ denotes the arrival signal phase of the $l${th} component. We assume that $\alpha_l$'s are independent and uniformly distributed random variables between $0$ and $2\pi$, i.e., $\alpha_l\sim \mathcal{U}\big[0, 2\pi \big)$, $l=1,\cdots,L$. Note that the power gain $g_l$ in general is determined by  the distance-dependent signal attenuation and shadowing. It is also generally a function of the AoA $\theta_l$ since the effective aperture of the ULA varies with $\theta_l$.

 Note that by setting $L=1$, \eqref{eq:hn} reduces to the case with single-path transmission only, e.g., the line-of-sight (LOS) scenario. In this case, all $N$ antenna elements at the receiver have the same received signal power. In another special case when  $L\rightarrow \infty$, $h_i$ becomes a circularly symmetric complex Gaussian (CSCG) random variable, which leads to the well-known  Rayleigh fading channel model \cite{goldsmith2005wireless}.

The received baseband signal  at the BS is given by
\begin{align}\label{eq:yt}
\mathbf y=\sqrt{P_t}\mathbf h s + \mathbf n,
\end{align}
where $P_t$ is the transmitted signal power; $s$ is the information-bearing   signal of the user with unit  power; $\mathbf h\in \mathbb{C}^{N\times 1}$ denotes the vector consisting of the complex-valued coefficients of the single-input multiple-output  (SIMO) channel, i.e., $\mathbf h=\left[h_1,\cdots, h_N \right]^T$;  and $\mathbf n\in \mathbb{C}^{N\times 1}$ stands for the additive noise vector with components modeled by independent and identically distributed (i.i.d.) zero-mean CSCG random variables with variance $\sigma^2$, i.e., $\mathbf n \sim \mathcal{CN}(\mathbf 0, \sigma^2 \mathbf I_N)$. 

From \eqref{eq:hn} and \eqref{eq:yt}, it follows  that the $l${th} path component arrives at all the $N$ receiving antennas with equal power $P_tg_l$, $\forall l=1,\cdots,L$. The total power collected by the $N$-element antenna array due to the $l${th} path  component is thus  given by $NP_tg_l$.

\section{EM Lens Embedded Antenna Array}\label{sec:proposed}

  In this section, we investigate further our proposed ELA design for antenna arrays equipped with a front layer of dielectric lens,  as shown in  Fig.~\ref{F:proposed}. First, we characterize the power distribution at receiving antennas after the EM lens filtering as a function of the AoA. Then we model the resulting SIMO channel with an ELA at the receiver.
 \subsection{Power Distribution of EM Lens}
\begin{figure}
\centering
\includegraphics[width=2.8in, height=2in]{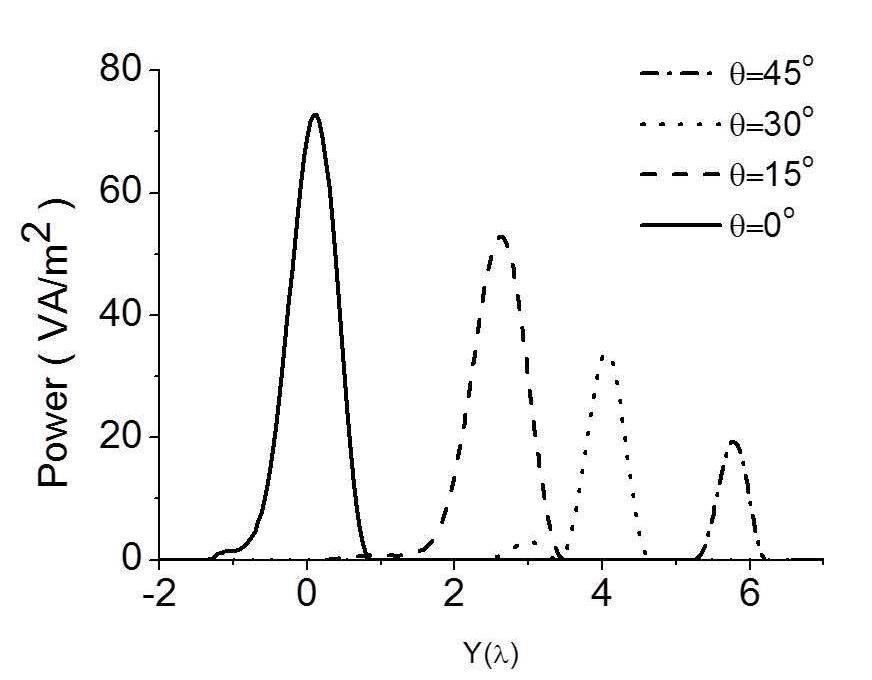}
\caption{Power distribution of an EM lens with different  AoA values $\theta$ (reproduced based on Fig. 3 in \cite{376}).\vspace{-3ex}}
\label{F:RFPowerdistribution}
\end{figure}

  As illustrated in Fig.~\ref{F:proposed}, an EM lens has two main functions, which are energy focusing and path separation in space, respectively. To specify these functions, we denote $p(y;\theta)$ as the power density function along the y-axis at the plane of the receiving antenna array, which is activated by an incident plane wave with AoA of $\theta\in [-\frac{\Theta}{2}, \frac{\Theta}{2}]$ upon  the EM lens. Note that $p(y;\theta)$ is defined over $-\frac{D}{2}\leq y \leq \frac{D}{2}$, where $D$ is the aperture diameter of the EM lens, whose center is placed at $y=0$ as shown in Fig.~\ref{F:proposed}. Also note that for a plane wave with AoA $\theta$, the total power collected by the EM lens, denoted by $P_{\theta}$, is given by
\begin{align}
P_{\theta}=\int_{-D/2}^{D/2} p(y;\theta)dy.
  \end{align}
   A practical example of the power distribution function $p(y;\theta)$ of an EM lens is shown in Fig.~\ref{F:RFPowerdistribution}, from which we can observe that: 1) For each incident plane wave, a bell shape power density function is resulted, which demonstrates the energy focusing capability of the EM lens; 2) As $\theta$ increases, the peak power locations shift to the right along y-axis. This implies that the arriving multipath signals with different  AoA values can be spatially separated after passing through the EM lens; and 3) The total power $P_{\theta}$ collected by the EM lens decreases with increasing  $|\theta|$.\footnote{This is due to the fact that the effective aperture of the EM lens is in general  proportional to $\cos \theta$, which decreases with increasing  $|\theta|$.}

For convenience, we define the \emph{normalized} power density function as
  \begin{align}
 f(y;\theta)\triangleq \frac{p(y;\theta)}{P_{\theta}},
 \end{align}
 thus we have $\int_{-D/2}^{D/2} f(y;\theta)dy=1$, $\forall \theta \in [-\frac{\Theta}{2}, \frac{\Theta}{2}]$. From  the results  shown in Fig.~\ref{F:RFPowerdistribution},  it is observed that the  power density $p(y;\theta)$ for a given $\theta$  can be coarsely approximated by a Gaussian distribution. Therefore, in the rest of this paper, we assume the normalized power density $f(y;\theta)$ as a Gaussian distribution function with mean $\bar y_{\theta}$ and variance $V_{\theta}$, which specify the peak power location and average  power spread, respectively, for an incident wave with AoA of $\theta$. Thus, we have
 \begin{align}\label{eq:fytheta}
 f(y; \theta)=\frac{1}{\sqrt{2\pi V_{\theta}}}e^{-\frac{(y-\bar y_{\theta})^2}{2V_{\theta}}},
 \end{align}
 where $\bar y_{\theta}$ is given by
 \begin{align}\label{eq:xpeak}
 \bar y_{\theta}=\Big(\frac{\theta}{90}\Big)\Big(\frac{D}{2}\Big), \ \theta\in \Big[-\frac{\Theta}{2},\ \frac{\Theta}{2}\Big].
 \end{align}
 We further assume for simplicity that $V_{\theta}=V$, $\forall \theta\in [-\frac{\Theta}{2}, \frac{\Theta}{2}]$. Note that in \eqref{eq:fytheta}, we have neglected the boundary effect by letting $D\rightarrow \infty$ in order to have $\int_{-\infty}^{\infty} f(y;\theta)dy=1$. Furthermore, we define $\Delta$ as the $90\%$ power beamwidth, i.e., with a distance $\Delta/2$ away from the center of the  power distribution $\bar y_{\theta}$, the power level drops by $90\%$ off the peak value at the center. A simple calculation with Gaussian distribution reveals that $\Delta^2=18.42 V$.  Practically, $\Delta$ and $V$ both increase with the aperture of the EM lens, since a larger-size EM lens generally has a wider range of the focal area.

\subsection{Channel Model of ELA}
With the normalized power density  $f(y;\theta)$ given in \eqref{eq:fytheta}, we are now ready to derive the new channel coefficients for the SIMO system introduced in Section~\ref{sec:model} with an ELA at the receiver. We assume the EM lens and the antenna array are appropriately designed and installed so that for the $l${th} path component with AoA $\theta_l\in [ -\frac{\Theta}{2},\ \frac{\Theta}{2}]$, $l=1,\cdots,L$, we have: 1) the power $P_{\theta_l}$ collected by the EM lens  is equal  to that  received by the antenna array without EM lens, i.e., $P_{\theta_l}=NP_tg_l$; and 2) the power received by the $i${th} antenna element with EM lens due to the $l${th} component, denoted as $\tilde{p}_{il}$, is obtained by integrating the power density function $p (y;\theta_l)$ as
\refstepcounter{equation}\label{eq:ptdil}
\begin{align}
\tilde{p}_{il}&=\int_{y_i-\frac{d}{2}}^{y_i+\frac{d}{2}} p(y;\theta_l)dy =P_{\theta_l}\underbrace{\int_{y_i-\frac{d}{2}}^{y_i+\frac{d}{2}} f(y;\theta_l)dy}_{\triangleq \beta_{il}} \notag \\
&=(NP_tg_l)\beta_{il}, \ i=2,\cdots, N-1, \tag{\arabic{equation}a}\\
\tilde{p}_{1l}&=(NP_tg_l)\beta_{1l}, \quad \tilde{p}_{Nl}=(NP_tg_l)\beta_{Nl}, \tag{\arabic{equation}b}
\end{align}
where $\beta_{1l}\triangleq \int_{-\infty}^{y_1+\frac{d}{2}} f(y;\theta_l)dy$, $\beta_{Nl}\triangleq\int_{y_N-\frac{d}{2}}^{\infty} f(y;\theta_l)dy$, and $y_i$ denotes the location of the $i${th} antenna element, which is given by \eqref{eq:yi}.
Note that $\beta_{il}$ represents the fraction of the power received by the $i${th} antenna element due to the $l${th} component. We thus have $\sum_{i=1}^N \beta_{il}=1$, $\forall l=1,\cdots,L$.
With $\tilde p_{il}$ defined in \eqref{eq:ptdil}, the  channel coefficient in \eqref{eq:hn} for the $i${th} antenna with the EM lens is modified as
\begin{equation}\label{eq:hntd}
\begin{aligned}
\tilde{h}_i=\sum_{l=1}^L \sqrt{Ng_l \beta_{il}}e^{j(\alpha_l+\frac{2\pi}{\lambda} i d \sin \theta_l+\eta_{il})}, \ i=1,\cdots,N,
\end{aligned}
\end{equation}
where $\eta_{il}$ is the phase change due to the EM lens, which is assumed to be independently distributed  over both $i$ and $l$ as $\eta_{il}\sim \mathcal{U}\big[0,\ 2\pi\big)$.
Let $\htd=\big[\tilde h_1,\cdots, \tilde h_N \big]^T$. Then the received signal  $\tilde{\mathbf y}$ at the BS with an ELA is given by
\begin{align}\label{eq:ytnew}
\tilde{\mathbf y}=\sqrt{P_t} \htd s + \mathbf n,
\end{align}
where $P_t$, $s$ and $\mathbf n$ have been defined in \eqref{eq:yt}.

\begin{figure}
\centering
\includegraphics[scale=0.6]{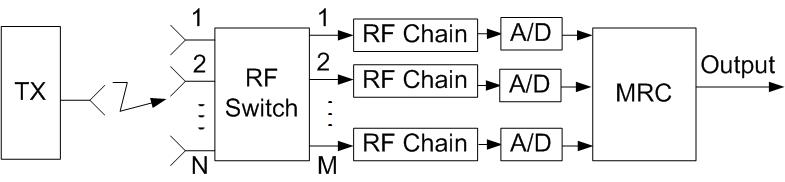}
\caption{Antenna selection.}
\label{F:AntennaSelection}
\end{figure}

\section{Antenna Selection and Performance Analysis}\label{sec:major}
In this section, we compare the performance for the SIMO system with versus without the EM lens under
the antenna selection framework.
\subsection{Antenna Selection}
As the number of receiving antennas $N$ becomes too large, it is costly in terms of both hardware implementation and energy consumption to make all antennas operate  at the same time. A practical  low-cost solution is thus  \emph{antenna selection} (AS) \cite{369}, where the ``best'' subset of $M$ out of $N$, $M\leq N$, receiving antennas are selected for processing the received signal, as shown in Fig.~\ref{F:AntennaSelection}. AS  reduces the number of required RF chains significantly from $N$ to $M$ if $M$ is much smaller than $N$. For SIMO systems, the optimal combining scheme for the output signals of the $M$ selected antennas is known to be maximal-ratio combining (MRC), with the optimal weights given as the conjugated complex coefficients of the corresponding channels \cite{goldsmith2005wireless}. As a result, the combiner output SNR is equal  to the sum of individual branches' SNRs.
For example, for  the SIMO system given in \eqref{eq:yt} without the EM lens, we denote the received SNR of the $i${th} antenna as
\begin{align}
&\gamma_i=\frac{P_t}{\sigma^2} |h_i|^2, \ i=1,\cdots,N. \label{eq:gamman}
\end{align}
Then, for AS with a given $M\leq N$, it follows that the maximum combiner output SNR, denoted as $\gamma_{(M)}$, is achieved by combining the $M$ branches with the highest SNRs via MRC, which yields
\begin{align}
\gamma_{(M)}=\sum_{i=1}^M \gamma_{[i]}, \label{eq:gammaM}
\end{align}
where $[\cdot]$ denotes a permutation such that $\gamma_{[1]}\geq \gamma_{[2]}\geq \cdots \geq \gamma_{[N]}$. The achievable rate in bits/sec/Hz (bps/Hz) for the  SIMO channel without the EM lens is then given by \cite{goldsmith2005wireless}
\begin{align}
R_{M}=\log_2\Big(1+\gamma_{(M)}\Big). \label{eq:CM}
\end{align}

Similarly, for the SIMO system given in \eqref{eq:ytnew} with the EM lens, we denote  the received SNR of the $i${th} antenna as $\gammatd_i$,  the output SNR by combining  the $M$ strongest  antenna signals via MRC as $\gammatd_{(M)}$, and  the achievable rate as $\Rtd_{M}$, which are given by
\begin{align}
&\gammatd_i=\frac{P_t}{\sigma^2} |\tilde h_i|^2, \ i=1,\cdots,N, \label{eq:gammatdn} \\
&\gammatd_{(M)}=\sum_{i=1}^M \gammatd_{[i]}, \ 1\leq M\leq N, \label{eq:gammatdM}\\
&\Rtd_{M}=\log_2\Big(1+\gammatd_{(M)}\Big).\label{eq:CtdM}
\end{align}
Note that if $M=N$, for both SIMO systems with or without the EM lens, MRC is indeed capacity optimal \cite{goldsmith2005wireless}, i.e., $R_{N}=C_N$, $\Rtd_{N}=\Ctd_N$, where $C_N$ and $\Ctd_N$ denote the capacity of the SIMO channels given in \eqref{eq:yt} and \eqref{eq:ytnew}, respectively.


\subsection{Rate Comparison}\label{sec:LOS}
Next, we compare the achievable rates $R_{M}$ and $\Rtd_{M}$ by AS with given $M\leq N$ for the SIMO systems with versus without the EM lens. For simplicity, we consider the scenario with one  single propagation  path only (i.e., $L=1$), say the $l$th path, which may be the LOS path in an outdoor environment for example. From  \eqref{eq:hn} and \eqref{eq:hntd}, we then have $|h_i|^2=g_l$, $\forall i$, and $|\tilde h_i|^2=Ng_l\beta_{il}$, with $\beta_{il}$ defined in \eqref{eq:ptdil}. In the following, we omit the path index $l$ for  brevity. Then the SNRs received by each antenna given in \eqref{eq:gamman} and \eqref{eq:gammatdn} for the single-path case reduce to
\begin{align}
&\gamma_i=\frac{P_tg}{\sigma^2}, \ i=1,\cdots, N, \label{eq:gammanLOS}\\
&\tilde \gamma_i=\frac{P_tg}{\sigma^2}N\beta_i,\  i=1,\cdots, N \label{eq:gammatdnLOS}.
\end{align}

Since $\sum_{i=1}^N \beta_{i}=1$, if all the $N$ antennas are used, i.e. $M=N$, from \eqref{eq:gammaM} and \eqref{eq:gammatdM}, we have $\gamma_{(N)}=\gammatd_{(N)}=\frac{P_tg}{\sigma^2}N$. As a result, with single-path transmission only,  there is no rate/capacity improvement for the system with over  without EM lens, i.e., $\Rtd_N=R_N$, and $\Ctd_N=C_N$. However, if $M<N$, we show in the following  via \emph{majorization theory} that a strictly positive rate gain can be achieved with the EM lens.


\begin{definition}\label{def:majorization}
(\textbf{Majorization}\cite{372}) Given $\mathbf x, \mathbf y \in \mathbb{R}^N$, $\mathbf x$ is majorized by $\mathbf y$, denoted as $\mathbf x \prec \mathbf y$, if
\begin{align}
&\sum_{i=1}^M x_{[i]}\leq \sum_{i=1}^M y_{[i]}, \  M=1,\cdots, N-1, \label{eq:major}\\
&\sum_{i=1}^N x_{[i]}= \sum_{i=1}^N y_{[i]}.
\end{align}
\end{definition}

Intuitively, $\mathbf x \prec \mathbf y$ indicates that the elements in $\mathbf x$ are ``less spread out'' than those  in $\mathbf y$. A simple result from majorization theory is that a vector with equal components is  majorized by another vector that has the same elementwise sum, as given by the following lemma.
\begin{lemma}\label{lemma:majorization}
Given $\mathbf{\tilde x}=[\tilde x_1, \cdots, \tilde x_N]^T$ and  $\mathbf x =[x_1,\cdots,x_N]^T$ with $x_i=\sum_{k=1}^N \tilde{x}_k/N$, $\forall i=1,\cdots, N$, then we have
\[
\mathbf x \prec \mathbf{\tilde x}.
\]
\end{lemma}

Let \gammabd \ and \gammatdbd \  be the two vectors consisting of the branch SNRs given in \eqref{eq:gammanLOS} and \eqref{eq:gammatdnLOS}, respectively. Since $\sum_{i=1}^N \beta_i=1$, it then follows from  Lemma~\ref{lemma:majorization} that \gammabd $\prec$ \gammatdbd. Thus, we have the following proposition.
\begin{proposition}\label{proposition}
For the single-path transmission case with $L=1$, we have
\[
\Rtd_{M}\geq R_{M}, \ 1\leq M<N,
\]
where the strict inequality holds if in \gammatdbd, all the elements $\gammatd_i$'s given in \eqref{eq:gammatdnLOS} are non-equal.
 \end{proposition}

Proposition~\ref{proposition} indicates that with AS, the rate of the SIMO system with the EM lens is always larger than that without the EM lens given non-identical antenna SNRs after the EM lens filtering (which is practically always the case; see Fig.~\ref{F:RFfileddistribution})  under the single-path setup.


\section{Numerical Results}\label{sec:simulation}
In this section, we compare the achievable rates for the SIMO systems with versus without the EM lens by simulations. The BS is assumed to be equipped with an $20$-element ULA ($N=20$) with half-wavelength separation between adjacent antennas ($d=\lambda/2$).  The normalized power density  $f(y; \theta)$ of the EM lens is given by \eqref{eq:fytheta}, with the $90\%$ power beamwidth set as $\Delta=3\lambda$. The AoA values $\theta_l$, $l=1,\cdots,L$, are assumed to be independent and uniformly distributed  between $-\frac{\Theta}{2}$ and $\frac{\Theta}{2}$ $\big(\theta_l\sim \mathcal{U} [-\frac{\Theta}{2}, \frac{\Theta}{2}]\big)$, where $\Theta$ is the angular spread. For the $l$th path with given $\theta_l$, by taking into account the effective aperture of the ULA, we assume that the power gain $g_l$ is proportional to $\cos \theta_l$, i.e., $g_l\propto \cos \theta_l$, where the proportional constant depends on the path-loss and shadowing. The power gain $g_l$'s are normalized so that $\frac{P_t}{\sigma^2}\sum_{l=1}^L g_l=\Gamma$, where $\Gamma$ is a parameter indicating the average received SNR level for both the cases with and without the EM lens at the receiver. Without loss of generality, we assume in the following that $\frac{P_t}{\sigma^2}=1$, so that $g_l= \frac{\Gamma \cos \theta_l}{\sum_{k=1}^L \cos \theta_k}$, $l=1,\cdots,L$.  

\subsection{Rate Comparison in Single-Path Environment}
\begin{figure}
\centering
\includegraphics[scale=0.26]{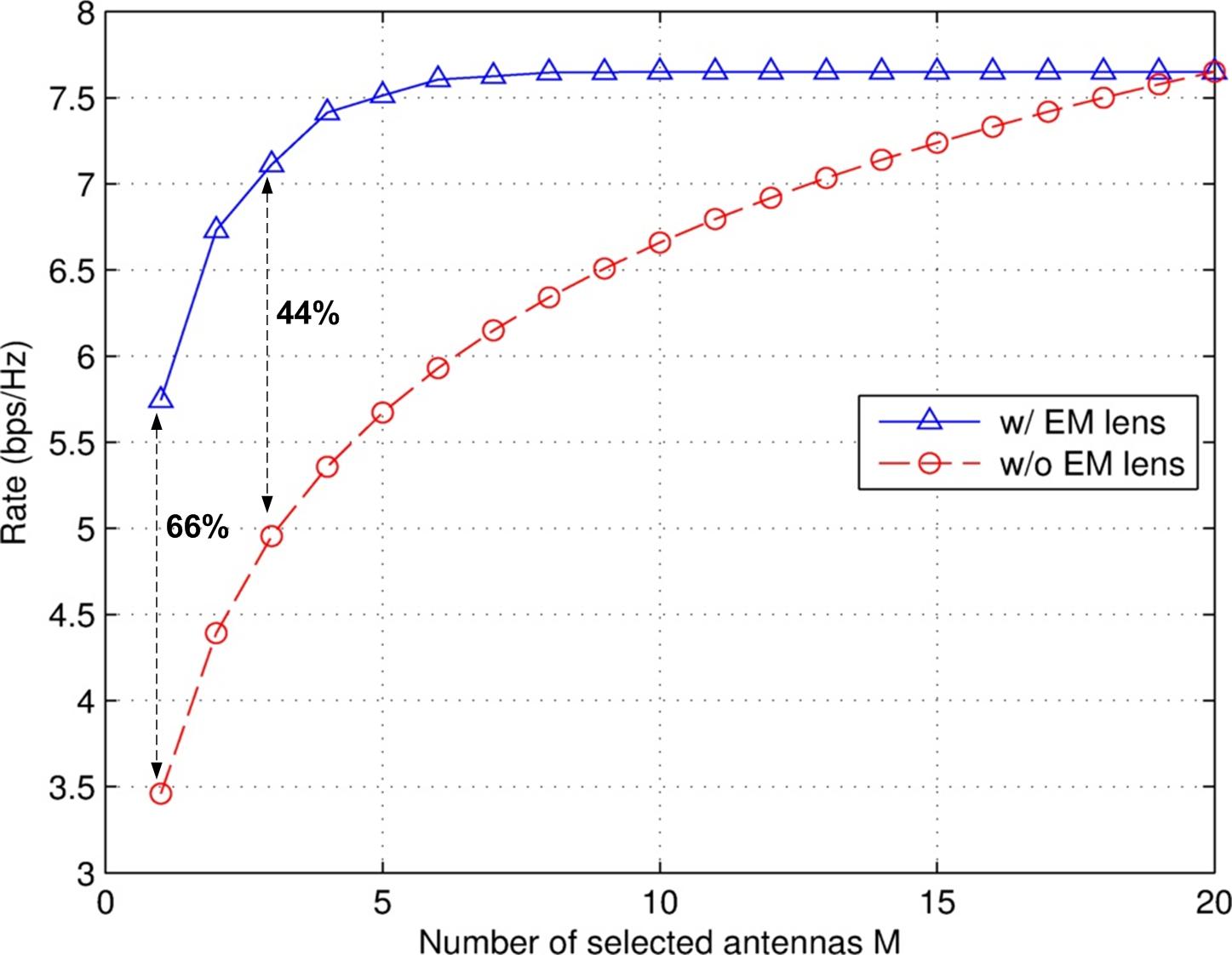}
\caption{Rate versus number of selected antennas in single-path environment with $\Gamma=10$ dB.\vspace{-2ex}}
\label{F:CapacityvsMSNR10N20LOS}
\end{figure}

First, we consider the case  with single-path transmission, i.e., $L=1$ in \eqref{eq:hn}. In Fig.~\ref{F:CapacityvsMSNR10N20LOS}, the achievable rate versus the number of selected antennas for AS, $M$, is shown for $\Gamma=10$ dB. It is observed that the SIMO system with the EM lens strictly outperforms that without the EM lens for all values of $M<N$, while the two systems achieve the same rate (capacity) when $M=N$. This is in accordance with the analytical result given in Proposition~\ref{proposition}. It is also observed that the rate gain is more pronounced for smaller $M$. For example, with $M=1$ or $M=3$, an $66\%$ or $44\%$ rate gain is achievable. Moreover, in order to achieve the rate within $99\%$ of the SIMO channel capacity, $19$ antennas need to be selected for AS without the EM lens, while this number is significantly reduced to $6$ in the case with EM lens, based on the results in Fig.~\ref{F:CapacityvsMSNR10N20LOS}.



\subsection{Rate Comparison in Multipath Environment}
\begin{figure}
\centering
\includegraphics[scale=0.19]{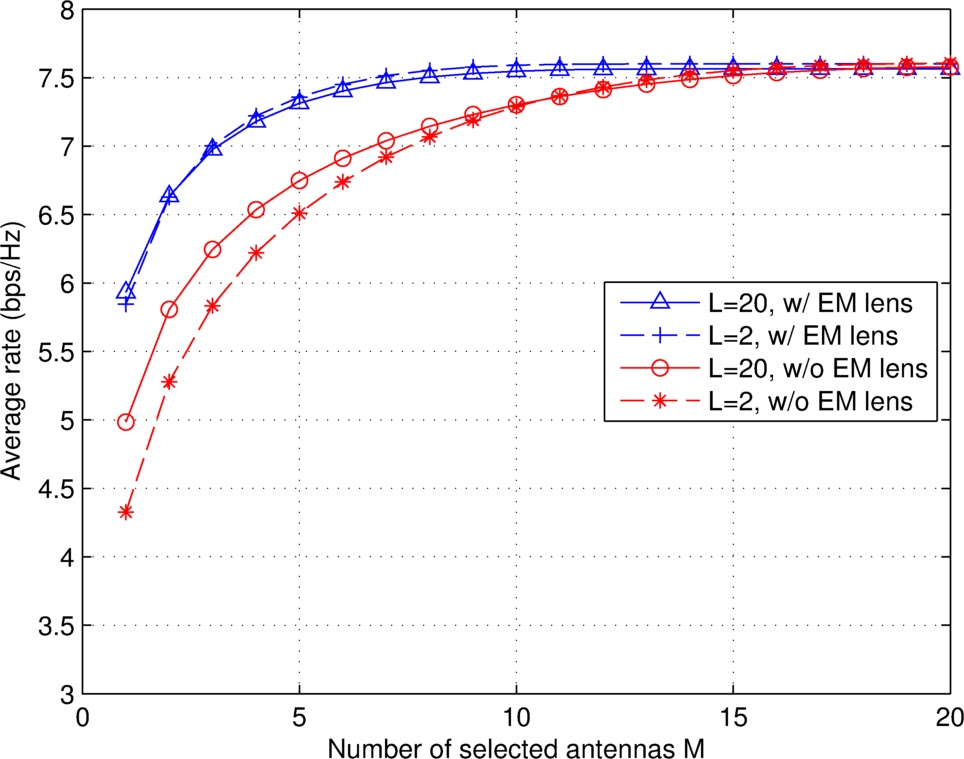}
\caption{Rate versus number of selected antennas in multipath environment with $L=2$ or $L=20$, $\Gamma=10$ dB, and $\Theta=60^{\circ}$.\vspace{-3ex}}
\label{F:CapacityvsMSNR10N20K0Theta60}
\end{figure}

Next, we consider a multipath environment with $L>1$ paths. With $L=2$ or $20$, we plot in Fig.~\ref{F:CapacityvsMSNR10N20K0Theta60} the achievable rates averaged over $1000$ random channel realizations with $\Gamma=10$ dB and $\Theta=60^{\circ}$ (which may correspond to a practical BS antenna array for one of the six equally-covered sectors in a cell). Significant rate gains are observed for both $L$ values considered at relatively small $M$ for the SIMO system with versus  without the EM lens. It is also observed that the achievable rate of the SIMO system with the EM lens is insensitive to $L$; however, increasing $L$ helps improve the rate of the system without EM lens.  This can be explained as follows. For the system without EM lens, the received power at each receiving antenna is due to the superposition of $L$ independent  multipath components that have equal power over all antennas. As a result, larger $L$ leads to more  significant power fluctuations  across antennas, which makes AS more effective. In contrast, for the system with EM lens, the power of each multipath component is mainly distributed over a certain subset of receiving antennas, which are different for multipaths with different AoAs; thus, the diversity effect by increasing $L$ is less notable.


\section{Conclusion and Future Work}\label{sec:conclusion}
In this paper, we have proposed a novel antenna
 system design for massive MIMO, where an EM lens is
 deployed together with the large antenna array, termed
 electromagnetic lens antenna (ELA). An ELA has been shown
 to offer two main benefits, namely spatial multipath
 separation and energy focusing. Under a simplified
 single-user uplink transmission setup, we have characterized
 the power distribution of the EM lens, and thereby
 established a new channel model for the SIMO system
 with a receiver ELA. Furthermore, we have demonstrated the
 significant rate gains of the proposed ELA system over
 conventional antenna arrays assuming the same antenna
 selection scheme applied at the receiver.

Several important issues remain to be addressed for the proposed ELA system in our future work: 
 \begin{itemize}
 \item{\emph{Power Distribution.} The performance of an ELA critically depends on the energy focusing and path separation capabilities of the EM lens. Therefore, a more accurate  characterization for the  power density function of the EM lens is desirable.} 
 \item{\emph{Channel Knowledge.} The performance analysis in this paper  assumes perfect channel knowledge. It is thus  necessary to study the system for a more practical  scenario  with imperfect channel estimation.} 
 \item{\emph{Multiuser System.} Investigating the proposed system in the multi-user setup is  promising. In particular, the spatial multipath separation capability of the proposed ELA system can be further exploited for   interference rejection, since the arriving signals from  different users generally have different AoAs. This  provides a possible solution to  the ``pilot contamination'' problem \cite{374} in massive MIMO.}
 \item{\emph{Downlink Transmission.} It is interesting  to study the downlink transmissions with an ELA at the BS transmitter, where the celebrated uplink-downlink duality via channel reciprocity is worth revisiting.}
 \end{itemize}

\bibliographystyle{IEEEtran}
\bibliography{IEEEabrv,IEEEfull}

\end{document}